\documentclass[twocolumn,showpacs,preprintnumbers,amsmath,amssymb,nofootinbib]{revtex4-1}

\usepackage{graphicx}
\usepackage{dcolumn}
\usepackage{bm}
\usepackage{amsfonts}

\def\bequ{\begin{equation}}
\def\eequ{\end{equation}}
\def\be{\begin{equation}}
\def\ee{\end{equation}}

\begin{document}


\title{Superradiant instabilities in a
$D$-dimensional small \\ Reissner-Nordstr\"om-Anti-de Sitter black hole }

\author{Mengjie Wang}
\email{mengjie.wang@ua.pt}
\author{Carlos Herdeiro}
\email{herdeiro@ua.pt}
\affiliation{\vspace{2mm}Departamento de F\'\i sica da Universidade de Aveiro and I3N \\
Campus de Santiago, 3810-183 Aveiro, Portugal \vspace{1mm}}%

\date{\today}

\begin{abstract}
We investigate the superradiant instability for a charged scalar field in a $D$-dimensional small Reissner-Nordstr\"om-Anti-de Sitter (RN-AdS) black hole. Firstly, we solve the charged Klein-Gordon equation analytically by a matching method.
We show that the general $D$-dimensional quasinormal frequencies depend on the relation between the angular momentum quantum number, $\ell$, and $D$. When $\ell$ is a (non-negative) integer multiple of $D-3$, i.e $\ell=p(D-3)$, we find an analytical quasinormal frequency formula adding a purely imaginary correction to the AdS normal frequencies. This is the case  for all $\ell$ modes in $D=4$. For general $D$ there are two more cases: i) when $\ell$ obeys $\ell=(p+\frac{1}{2})(D-3)$, which can occur for odd $D$, we observe that
the matching method fails, since the near and far region solutions have different functional dependences; ii) for all other cases, the analytical quasinormal frequency formula gives a complex correction to the AdS normal frequencies. Secondly,  we perform a numerical calculation which confirms the analytical formulas obtained with the matching method and allows us to explore the case where that method failed. In the latter case, as in the former, we verify that all $\ell$ modes for all $D$ may become superradiant,  which contradicts a previous claim in the literature.
\end{abstract}

\maketitle


\section{Introduction}

In black hole (BH) physics, the phenomenon of superradiance~\cite{zeldovich1,zeldovich2} is a wave analogue of the Penrose process~\cite{Penrose:1969pc}, in the sense of allowing energy extraction from a rotating BH. For a bosonic field wave with the form $e^{-i\omega t+im\phi}$ impinging on a rotating BH with angular velocity $\Omega$, it turns out that the scattered wave is amplified when $\omega < m \Omega$, which is consistent with both the first and second law of BH thermodynamics. It follows that, by placing a reflecting mirror around a rotating BH, the system becomes unstable. The role of the mirror is to feed back to the BH the amplified scattered wave, as to recurrently extract rotational energy. Then, the wave bounces back and forth between the mirror and the BH until radiation pressure destroys the mirror. This process was dubbed as \textit{black hole bomb} by Press and Teukolsky~\cite{Press:1972zz}. Such BH mirror system has been investigated in the linear regime in both the frequency~\cite{Cardoso:2004nk,Lee:2011ez} and time domain~\cite{Dolan:2012yt}, to obtain the time scale of the instability. Actually, the reflecting mirror is not necessarily artificial, and it has several realizations in nature. One realization is the field's mass. For a massive bosonic field with mass $m_0$ satisfying the bound state condition $\omega < m_0$, the mass term can work as a mirror~\cite{Detweiler:1980uk,Dolan:2007mj,Rosa:2009ei,Hod:2009cp,Witek:2012tr,Rosa:2012uz,Pani:2012vp,Hod:2012zza}. Another realization is Anti-de Sitter (AdS) asymptotics, which provides an infinite potential at infinity, binding superradiant modes~\cite{Cardoso:2004hs,Cardoso:2006wa,Aliev:2008yk,Dias:2011at,Li:2012rx,Uchikata:2011zz,Cardoso:2013pza}. A considerable interest has been recently devoted to non-linear studies of superradiant scattering and instability, using numerical methods \cite{Cardoso:2012qm,Yoshino:2012kn,Witek:2012tr,Yoshino:2013ofa,East:2013mfa,Okawa:2014nda}.

The superradiance phenomenon does not exist only for rotating BHs; it also exists for charged BHs. For a charged bosonic field mode with frequency $\omega$ and charge $e$ impinging on a charged BH with electric potential at the horizon $\Phi_H$, Coulomb energy and BH charge can be extracted when $\omega<e\Phi_H$ \cite{Bekenstein:1973ur}.  As argued in~\cite{Furuhashi:2004jk,Hod:2013eea}, however, the superradiance condition and the bound state condition for a charged massive scalar field in an asymptotically flat charged BH cannot be satisfied simultaneously. Thus, a charged BH bomb requires AdS asymptotics or a mirror-like boundary condition, i.e. considering a BH in a cavity. The latter case has been recently investigated in the linear regime in both the frequency \cite{Herdeiro:2013pia,Hod:2013fvl} and time domain \cite{Degollado:2013bha}.

The superradiant instability for a charged scalar field in a charged AdS BH has been explored in~\cite{Uchikata:2011zz} for four dimensions and in~\cite{Aliev:2008yk,Dias:2011tj} for five dimensions. A complete analysis, however, in a $D$-dimensional charged AdS BH is still lacking. Presenting such complete study is the goal of this paper. As a spin off we shall clarify a claim, made in~\cite{Aliev:2008yk}, that, in $D=5$,  a subset of field modes - those with odd angular momentum quantum number, $\ell$ - do not develop the superradiant instability. We shall show otherwise: that in fact \textit{all} $\ell$ modes in \textit{all} dimensions can develop the superradiant instability.

Besides the superradiant instability due to the presence of a minimally coupled charged scalar field, other studies on stability have been considered for charged AdS BHs. It was first observed in \cite{Gubser:2000ec,Gubser:2000mm} that for BHs obtained within $\mathcal{N}=8$ supergravity in $D=4$, which is a theory containing 4 $U(1)$ gauge fields and 3 real scalar fields \textit{non-minimally} coupled to the Maxwell fields, \textit{large} RN-AdS BHs are dynamically  unstable due to the existence of a tachyonic mode in the scalar field perturbations. On the other hand, in purely Einstein-Maxwell theory in $D$ dimensions, Reissner-Nordstr\"om-AdS (RN-AdS) BHs have been argued to be stable against gravitational perturbations~\cite{Konoplya:2008rq}. Thus the existence of scalar fields with some coupling to Maxwell fields is central to the instability of~\cite{Gubser:2000ec,Gubser:2000mm}. A qualitatively different instability   of RN-AdS BHs has been discussed in the context of holographic superconductors. It occurs in the presence of a massive scalar field that may or may not be charged and leads to the formation of scalar hair around the BH \cite{Hartnoll:2008kx}. Unlike the superradiant instability, this other instability occurs even if the scalar field is uncharged. In that case the, say, $D=4$ RN-AdS BH should be nearly extremal, which means its near horizon geometry has a two dimensional AdS factor; moreover, the scalar field should have a tachyonic mass above the Breitenlohner-Freedman  bound \cite{Breitenlohner:1982bm,Breitenlohner:1982jf} of the four dimensional AdS space, but below the Breitenlohner-Freedman bound of the two dimensional AdS factor in the near horizon geometry of the BH. This is why the scalar field gives an instability of the BH geometry, without being an instability of the four dimensional AdS space.

In order to investigate the superradiant instability of a charged, massive scalar field in a $D$-dimensional RN-AdS background we shall employ both an analytical and a numerical method. Firstly, we solve the charged Klein-Gordon equation using a matching method between a near (the BH) region solution and a far region solution, for small BHs, i.e. BHs obeying $r_+\ll L$, where $r_+$ and $L$ stand for the BH event horizon and the AdS radius. We thus obtain the analytical quasinormal frequency for small BHs by matching the near and far solutions in an intermediate region. We find that the relations between $\ell$ and $D$ play a central role in determining the analytical quasinormal frequency formula. When $\ell=(p+\frac{1}{2})(D-3)$, where $p$ is a non-negative integer, the matching method fails. The reason is that the near region solution and the far region solution have different functional dependence in terms of the radial coordinate, which makes matching impossible. Such difficulty in employing the matching method also occurs for extremal BHs~\cite{Rosa:2009ei}, where an alternative point matching method was used. For all other relations between $\ell$ and $D$, the matching method works and it may be used to show that the superradiance instability exists for all $\ell$ modes, in a region of the parameters space.

Then we solve the charged Klein-Gordon equation numerically both to check the analytical results and to explore the special case where the analytical method fails. We find good agreement between the analytical approximation and the numerical results. For the special case $\ell=(p+\frac{1}{2})(D-3)$, the numerical results show that the superradiant instability does exist.

The structure of this paper is organized as follows. In Section~\ref{seceq} we introduce the background geometry and the scalar field equation which will be explored in this paper. In Section~\ref{secmatching} we solve the scalar field equation analytically by the matching method and obtain an analytical quasinormal frequency formula. We analyze this formula for different relations between $\ell$ and $D$, and show the reason why the matching method fails for a special case, i.e. when $\ell=(p-\frac{1}{2})(D-3)$, in Section~\ref{result analysis}. To confirm our analytical results and to be able to investigate if there is a superradiant instability for that special case, we appeal to a numerical method to solve the Klein-Gordon equation in Section~\ref{numerical}.
Final remarks are presented in the last section.

\section{background and field equation}
\label{seceq}
We consider a $D$-dimensional Reissner-Nordstr\"om-Anti-de Sitter BH with the line element
\begin{equation}
ds^2=-f(r)dt^2+\dfrac{1}{f(r)}dr^2+r^2d\Omega^2_n \;,\label{metric}
\end{equation}
where $d\Omega^2_n$ is the metric on the unit $n$-sphere. In the following it will be convenient to use $n=D-2$ rather than $D$ to parameterize the space-time dimension. The metric function $f(r)$ takes the form
\begin{equation}
f(r)=1-\dfrac{\mu}{r^{n-1}}+\dfrac{q^2}{r^{2(n-1)}}+\dfrac{r^2}{L^2} \;,\label{metricfunc}
\end{equation}
where the parameters $\mu, q$ and $L$ are related with the BH mass $M$, charge $Q$ and cosmological constant $\Lambda$ through
\begin{equation}
\mu=\dfrac{16\pi GM}{nS_n}\;\;,q^2=\dfrac{8\pi GQ^2}{n(n-1)}\;\;,L^2=-\dfrac{n(n+1)}{2\Lambda}\;,\nonumber
\end{equation}
and the area of a unit $n$-sphere is $S_n=\frac{2\pi^{\frac{n+1}{2}}}{\Gamma(\frac{n+1}{2})}$.
The Hawking temperature is given by
\begin{equation}
T=\dfrac{1}{4\pi}\left[\frac{(n-1)\mu}{r_+^n}-\frac{2(n-1)q^2}{r_+^{2n-1}}+\frac{2r_+}{L^2}\right]\;,\label{hawkingtemp}
\end{equation}
where the event horizon $r_+$ is determined as the largest root of $f(r_+)=0$. For non-extremal BHs, we have $q<q_c$, where the critical charge $q_c$ corresponds to the maximal charge (i.e the charge of an  extremal BH) and is given by
\begin{equation}
q_c\equiv r_+^{n-1}\sqrt{1+\dfrac{n+1}{n-1}\left(\dfrac{r_+}{L}\right)^2}\;.\label{critialq}
\end{equation}
The electromagnetic potential of the charged BH is
\begin{equation}
A=\left(-\sqrt{\dfrac{n}{2(n-1)}}\dfrac{q}{r^{n-1}}+C\right)dt\;,\nonumber
\end{equation}
where the choice of the constant $C$ is a gauge choice. For instance, we should fix $C$ as
\begin{equation}
C=\sqrt{\dfrac{n}{2(n-1)}}\dfrac{q}{r_+^{n-1}}\;,\nonumber
\end{equation}
in order to have a vanishing electromagnetic potential at the event horizon, a choice used in the context of the AdS/CFT correspondence~\cite{Horowitz:2010gk}. As argued in the paper~\cite{Uchikata:2011zz}, however, this constant just shifts the real part of quasinormal frequency as Re$(\omega)\rightarrow$ Re$(\omega)+eC$ ($e$ is the field charge) without affecting the imaginary part of it. Therefore, as we are mostly interested in the superradiant instability, which is determined by the imaginary part of the quasinormal frequency, we take in the following $C=0$.

For a charged massive scalar field, the corresponding Klein-Gordon (K-G) equation can be written as
\begin{eqnarray}
\dfrac{1}{\sqrt{-g}}D_\mu\left[\sqrt{-g}g^{\mu\nu}D_\nu\right]\phi-m_0^2\phi=0 \;,\label{sacalarEq}
\end{eqnarray}
where $D_\mu=\partial_\mu -ieA_\mu$, $e$ and $m_0$ are the field charge and mass, respectively. The scalar field $\phi$ can be decomposed in terms of spherical scalar harmonics due to the spherical symmetry of the background
\begin{equation}
\phi={\rm e}^{-i\omega t}\rm{R}(r)\rm{Y}(\theta_i) \;,\label{phicecompose}
\end{equation}
where $\rm{Y}(\theta_i)$ is a scalar spherical harmonic on the $n$-sphere.
Substituting the metric in Eq.~\eqref{metric} and the field decomposition in Eq.~\eqref{phicecompose} into the K-G equation~\eqref{sacalarEq}, we have
\begin{eqnarray}
&&\dfrac{\Delta}{r^{n-2}}\dfrac{d}{dr}\left(\dfrac{\Delta}{r^{n-2}}\dfrac{d\rm{R}}{dr}\right)-\Big(\lambda+m_0^2r^2\Big)\Delta\rm{R} \nonumber\\
&&+r^{2n}\Big(\omega+eA_t\Big)^2 \rm{R}=0 \;,\label{RadialEq}
\end{eqnarray}
with
\begin{equation}
\Delta\equiv r^{2(n-1)}-\mu r^{n-1}+q^2+\dfrac{r^{2n}}{L^2}\;,\nonumber
\end{equation}
and the $n-$dimensional spherical harmonic eigenvalue $\lambda$ is given by
\begin{equation}
\lambda\equiv \ell(\ell+n-1)\;,\nonumber
\end{equation}
where $\ell$ is the angular momentum quantum number.

For numerical convenience, we may rewrite Eq.~\eqref{RadialEq} in terms of a new function $\rm{X}$
\begin{eqnarray}
f(r)^2\dfrac{d^2\rm X}{dr^2} + f(r)f'(r) \dfrac{d \rm X}{dr} + \left[(\omega+eA_t)^2 -f(r)\right.\nonumber\\\left. \left(\dfrac{\lambda}{r^2}+m_0^2+\dfrac{nf'(r)}{2r}+\dfrac{n(n-2)f(r)}{4r^2}\right)\right] \rm X=0\;,\label{RadialEq2}
\end{eqnarray}
where ${\rm X}\equiv r^{n/2} \rm R$.

In order to determine the quasinormal frequency, by solving Eq.~\eqref{RadialEq2}, one has to impose boundary conditions. Near the event horizon, we impose an ingoing boundary condition
\begin{equation}
{\rm X} \sim {\rm e}^{-i\left(\omega-\omega_0\right)r_\ast}\;,\label{boundaryingoing}
\end{equation}
where $\omega_0\equiv -eA_t(r_+)$ and the tortoise coordinate $r_\ast$ is defined by
\begin{equation}
\dfrac{dr_\ast}{dr}=\dfrac{1}{f(r)}\;.\nonumber
\end{equation}
At infinity, we impose a decaying boundary condition
\begin{equation}
{\rm X} \sim r^{-\frac{1}{2}(1+\sqrt{4m_0^2L^2+(n+1)^2})}\; ,\label{boundarydecaying}
\end{equation}
then
\begin{equation}
{\rm R} \sim r^{-\frac{1}{2}(n+1+\sqrt{4m_0^2L^2+(n+1)^2})}\;.\label{boundarydecaying2}
\end{equation}
For a massless field in $D$ dimensions, the radial function goes, asymptotically as ${\rm R}\sim 1/r^{D-1}$, as expected for a normalizable massless scalar perturbation in $D$-dimensional AdS space.

\section{Analytic calculations}
\label{secmatching}
In this section, we present the analytic calculations of quasinormal frequencies for a charged massive scalar field in a higher dimensional RN-AdS BH. Using a standard procedure, we shall divide the space outside the event horizon into two regions: the \textit{near region}, defined by the condition $r-r_+\ll1/\omega$, and the \textit{far region}, defined by the condition $r_+\ll r-r_+$. Then, in order to perform a matching of the two solutions we consider low frequency condition, obeying $r_+\ll1/\omega$, and match the two solutions in an  \textit{intermediate region} defined by $r_+\ll r-r_+\ll1/\omega$. In the following analysis we focus on small AdS BH $(r_+\ll L)$. This allows us to treat the frequencies for the RN-AdS BH as a perturbation of the AdS normal frequencies.

\subsection{Near region solution}
For the near region analysis, we rewrite Eq.~\eqref{RadialEq} as
\begin{eqnarray}
&&\left(n-1\right)^2\Delta\dfrac{d}{dx}\left(\Delta\dfrac{d {\rm R}}{dx}\right)-\Big(\lambda+m_0^2r^2\Big)\Delta {\rm R} \nonumber\\
&&+ r^{2n}\Big(\omega+eA_t\Big)^2 {\rm R}=0 \;,\label{NearEq1}
\end{eqnarray}
where $x\equiv r^{n-1}$. In the following we shall neglect the mass term in the first line of Eq.~\eqref{NearEq1}. This amounts to say that  $r\ll\frac{\ell}{m_0}$, which is obeyed if the Compton wave length of the scalar particles is much larger than BH horizon size and indeed becomes the near region condition if, moreover, the scalar particle Compton wave length is much smaller than the AdS radius. Observe that the condition  $r\ll\frac{\ell}{m_0}$ fails for $\ell=0$ modes, but it turns out that even in that case the analytical results we shall obtain are in good agreement with the numerical results for sufficiently small mass (cf. Sec.~\ref{numerical}). Under this approximation Eq.~\eqref{NearEq1} becomes
\begin{equation}
(n-1)^2\Delta\dfrac{d}{dx}\left(\Delta\dfrac{d{\rm R}}{dx}\right)-\lambda\Delta {\rm R} + r_+^{2n}\Big(\omega+eA_t\Big)^2 \rm R=0 \;.\label{NearEq2}
\end{equation}
For the convenience of analytic calculations, one can define a new dimensionless variable
\begin{equation}
z\equiv \dfrac{x-x_+}{x-x_-}\;,\nonumber
\end{equation}
with $x_+=r_+^{n-1}$ and $x_-=r_-^{n-1}$,
where $r_+$ and $r_-$ refer to the event horizon and the Cauchy horizon, respectively.
Then Eq.~\eqref{NearEq2} can be transformed into
\begin{equation}
z\dfrac{d}{dz}\left(z\dfrac{d{\rm R}}{dz}\right)+\left[\bar{\omega}^2-\dfrac{\lambda}{(n-1)^2}\dfrac{z}{(1-z)^2}\right] \rm R=0\;,\label{NearEq}
\end{equation}
with
\begin{equation}
\bar{\omega}\equiv \dfrac{x_+^{\frac{n}{n-1}}}{(n-1)(x_+ - x_-)}\left(\omega-\sqrt{\dfrac{n}{2(n-1)}}\dfrac{eq}{x_+}\right) \;.\label{superradiance}
\end{equation}
Observe that $\bar{\omega}<0$ for $\omega<\sqrt{\frac{n}{2(n-1)}}\frac{eq}{x_+}$. This will be shown below to correspond to the superradiant regime.

One may now obtain a solution for Eq.~\eqref{NearEq} with ingoing boundary condition in terms of a hypergeometric function:
\begin{equation}
{\rm R} \sim z^{-i\bar{\omega}}(1-z)^\alpha F(\alpha,\alpha-2i\bar{\omega},1-2i\bar{\omega};z) \;,\label{NearSol}
\end{equation}
where
\begin{equation}
\alpha\equiv 1+\dfrac{\ell}{n-1}\;.\label{alpha}
\end{equation}
In order to match the far region solution, one must expand the near region solution, Eq.~\eqref{NearSol}, at large $r$. To achieve this we take $z\rightarrow1$ limit and use the properties of the hypergeometric function~\cite{abramowitz+stegun}, then obtain
\begin{eqnarray}
{\rm R} \sim && \Gamma(1-2i\bar{\omega})\left[\dfrac{{\rm R}^{\rm near}_{1/r}}{r^{n-1+\ell}} +{\rm R}^{\rm near}_r r^{\ell}\right]
\;,\label{NearsolFar}
\end{eqnarray}
where
\begin{eqnarray}
{\rm R}^{\rm near}_{1/r} & \equiv & \dfrac{\Gamma(1-2\alpha) (r_+^{n-1} - r_-^{n-1})^\alpha}{\Gamma(1-\alpha)\Gamma(1-\alpha-2i\bar{\omega})} \ ,\nonumber \\
{\rm R}^{\rm near}_r &\equiv& \dfrac{\Gamma(2\alpha-1)(r_+^{n-1}-r_-^{n-1})^{1-\alpha}}{\Gamma(\alpha)\Gamma(\alpha-2i\bar{\omega})} \ .
\end{eqnarray}
Since the Gamma function has poles at negative integers, one observes that special care must be taken with the factor $\Gamma(1-2\alpha)/\Gamma(1-\alpha)$. Its analysis will play a role below.

\subsection{Far region solution}
In the far region, $r-r_+\gg r_+$, the BH effects can be neglected ($\mu\rightarrow0, q\rightarrow0$) so that
\begin{equation}
\Delta\simeq r^{2n} \left(\dfrac{1}{r^2}+\dfrac{1}{L^2}\right)\; .\nonumber
\end{equation}
Then Eq.~\eqref{RadialEq} becomes
\begin{eqnarray}
&&\left(1+\dfrac{r^2}{L^2}\right)\dfrac{d^2{\rm R}}{dr^2}+\left(\dfrac{n}{r}+\dfrac{(n+2)r}{L^2}\right)\dfrac{d {\rm R}}{dr}\nonumber\\
&&+\left[\dfrac{\omega^2}{1+\dfrac{r^2}{L^2}}-\left(\dfrac{\lambda}{r^2}+m_0^2\right)\right]\rm R=0\;.\label{fareq1}
\end{eqnarray}
Defining a new variable
\begin{equation}
y\equiv 1+\dfrac{r^2}{L^2}\;,\nonumber
\end{equation}
Eq.~\eqref{fareq1} becomes
\begin{eqnarray}
y\left(1-y\right)\dfrac{d^2{\rm R}}{dy^2}+\left(1-\dfrac{n+3}{2}y\right)\dfrac{d {\rm R}}{dy}\nonumber\\
-\left[\dfrac{\omega^2L^2}{4y}-\dfrac{m_0^2L^2}{4}+\dfrac{\lambda}{4(1-y)}\right]\rm R=0\;.\label{fareq2}
\end{eqnarray}
The above equation has a hypergeometric equation structure, which can be shown explicitly through the transformation
\begin{equation}
{\rm R}=y^{\frac{\omega L}{2}}(1-y)^{\frac{\ell}{2}}F(a,b;c;y) \;,\nonumber
\end{equation}
with parameters
\begin{eqnarray}
a&\equiv &\dfrac{n+1}{4}+\dfrac{\omega L}{2}+\dfrac{\ell}{2}+\dfrac{1}{2}\sqrt{m_0^2L^2+\left(\frac{n+1}{2}\right)^2}\;,\nonumber\\
b&\equiv &\dfrac{n+1}{4}+\dfrac{\omega L}{2}+\dfrac{\ell}{2}-\dfrac{1}{2}\sqrt{m_0^2L^2+\left(\frac{n+1}{2}\right)^2}\;,\nonumber\\
c&\equiv &1+\omega L\;.\label{hyperparameter}
\end{eqnarray}
Then considering a decaying boundary condition at infinity given in Eq.~\eqref{boundarydecaying2}, one finds a solution for Eq.~\eqref{fareq2} in the form
\begin{equation}
{\rm R}\thicksim (1-y)^{\frac{\ell}{2}}y^{\frac{\omega L}{2}-a} F(a,1+a-c;1+a-b;\frac{1}{y})\;.\label{farsolution}
\end{equation}
To achieve the small $r$ behavior for Eq.~\eqref{farsolution}, making the transformation $\frac{1}{y}\rightarrow1-y$ and using properties of the hypergeometric function~\cite{abramowitz+stegun}, we obtain
\begin{eqnarray}
{\rm R} \sim && \Gamma(1+a-b)\left[\dfrac{{\rm R}^{\rm far}_{1/r}}{r^{n-1+\ell}} +{\rm R}^{\rm far}_r r^{\ell}\right]   \;,\label{farsolution@near}\end{eqnarray}
where
\begin{eqnarray}
{\rm R}^{\rm far}_{1/r} & \equiv &  \dfrac{\Gamma(\ell+\frac{n-1}{2})L^{n-1+\ell}}{\Gamma(a)\Gamma(1+a-c)} \ ,\nonumber \\
{\rm R}^{\rm far}_r &\equiv& \dfrac{\Gamma(-\ell-\frac{n-1}{2})L^{-\ell}}{\Gamma(1-b)\Gamma(c-b)}\ .
\label{far_branches}
\end{eqnarray}
The solution~\eqref{farsolution@near} is for pure AdS. Regularity of the above solution at the origin $(r=0)$ of AdS requires \footnote{We remark that, alternatively, one can also demand $\Gamma(a)=0$, which gives the negative AdS spectrum. Without loss of generality, we only consider the positive spectrum in this paper.}
\begin{equation}
\Gamma(1+a-c)=\infty\ \qquad \ \Rightarrow \ \qquad 1+a-c=-N ,\nonumber
\end{equation}
which gives the discrete spectrum
\begin{equation}
\omega_{N} L=2N+\frac{n+1}{2}+\ell+\sqrt{m_0^2L^2+\left(\frac{n+1}{2}\right)^2}\;,\label{AdSfrequency}
\end{equation}
where $N$ is a non-negative integer.
Observe that the AdS frequencies remain real even for tachyonic modes when $0>m_0^2\ge -(n+1)^2/(4L^2)$. This is the well known Breitenlohner-Freedman bound already discussed in the introduction. In particular one may see that the bound is more negative for higher dimensional spaces. This is the reason why one may violate the bound for two dimensional AdS but obey it for four dimensional AdS, as discussed in the introduction.  For a massless scalar field the spectrum formula simplifies into
$\omega_{N} L=2N+(n+1)+\ell$.

When the BH effects are taken into account, a correction to the frequency (which can be complex) will be introduced
\begin{equation}
\omega=\omega_N+i\delta\;,
\end{equation}
where the real part of $\delta$ is used to describe the damping of the quasinormal modes.
Then, for small BHs, using the approximation $1/\Gamma(-N+\epsilon)\simeq (-1)^NN!\epsilon$ for small $\epsilon$, the first term appeared in the bracket of Eq.~\eqref{farsolution@near} becomes
\begin{eqnarray}
{\rm R}^{\rm far}_{1/r} = (-1)^{N+1} i\delta N! \dfrac{\Gamma(\ell+\frac{n-1}{2})L^{n+\ell}}{2\Gamma(a)} \;.\nonumber
\end{eqnarray}

Finally, observe that there appears to be extra poles in ${\rm R}_r^{\rm far}$, Eq.~\eqref{far_branches}, due to the Gamma function $\Gamma(-\ell-\frac{n-1}{2})$ for odd $n$. In the ${\rm R}_r^{\rm far}$ expression, however, due to \eqref{AdSfrequency}, $\Gamma(1-b)=\Gamma(-\ell-\frac{n-1}{2}-N)$, and thus canceling the former poles.

\subsection{Overlap region}
To match the near region solution~(\ref{NearsolFar}) and the far region solution~(\ref{farsolution@near}) in the intermediate region, we impose the matching condition ${\rm R}^{\rm near}_r{\rm R}^{\rm far}_{1/r}={\rm R}^{\rm far}_r{\rm R}^{\rm near}_{1/r}$, then $\delta$ can be obtained perturbatively
\begin{eqnarray}
\delta=&&2i\dfrac{(r_+^{n-1}-r_-^{n-1})^{2\alpha-1}}{L^{2\ell+n}} \dfrac{\Gamma(1-2\alpha)\Gamma(\alpha)}{\Gamma(2\alpha-1)\Gamma(1-\alpha)} \times \nonumber\\
&&\times \dfrac{(-1)^{N}}{N!} \dfrac{\Gamma(a)}{\Gamma(1-b)\Gamma(c-b)} \dfrac{\Gamma(-\ell-\frac{n-1}{2})}{\Gamma(\ell+\frac{n-1}{2})} \times \nonumber\\ &&\times \dfrac{\Gamma(\alpha-2i\bar{\omega})}{\Gamma(1-\alpha-2i\bar{\omega})}\;.\label{imaginarypart}
\end{eqnarray}
%

\section{Analytical result analysis}
\label{result analysis}
To analyze Eq.~\eqref{imaginarypart}, we shall simplify the Gamma functions therein. Firstly, the following combination, which is independent of the relation between $\ell$ and $n$, can be simplified as
\begin{eqnarray}
&&\dfrac{\Gamma(a)}{\Gamma(1-b)\Gamma(c-b)}\dfrac{\Gamma(-\ell-\frac{n-1}{2})}{\Gamma(\ell+\frac{n-1}{2})}\nonumber\\
&&=\dfrac{(-1)^N}{\Gamma(\ell+\frac{n-1}{2})} \dfrac{\Gamma(N+\frac{n+1}{2}+\ell+\sqrt{m_0^2L^2+(\frac{n+1}{2})^2})}{\Gamma(N+1+\sqrt{m_0^2L^2+(\frac{n+1}{2})^2})}\times\nonumber\\
&&\times\prod_{k=1}^{N}(\ell+\frac{n-1}{2}+k)\;.\nonumber
\end{eqnarray}
Then one has to consider the following cases separately, because the simplification for the other Gamma functions in Eq.~\eqref{imaginarypart} depends on the relation between $\ell$ and $n$.

\subsection{$\ell$ is an integer multiple of $(n-1)$}
For this case we can write $\ell=p(n-1)$, where $p$ is a non-negative integer.
Then, the corresponding Gamma functions in Eq.~\eqref{imaginarypart} can be simplified to
\begin{equation}
\dfrac{\Gamma(1-2\alpha)\Gamma(\alpha)}{\Gamma(2\alpha-1)\Gamma(1-\alpha)}=\dfrac{(-1)^{p+1}}{2}\dfrac{(p!)^2}{(2p)!(2p+1)!}\;,\nonumber
\end{equation}
\begin{equation}
\dfrac{\Gamma(\alpha-2i\bar{\omega})}{\Gamma(1-\alpha-2i\bar{\omega})}=(-1)^p 2i\bar{\omega} \prod_{k^{\prime}=1}^p(k^{\prime 2}+4\bar{\omega}^2)\;.\nonumber
\end{equation}
Therefore, Eq.~\eqref{imaginarypart} becomes
\begin{eqnarray}
&&\delta=-2\bar{\omega}\dfrac{(r_+^{n-1}-r_-^{n-1})^{1+\frac{2\ell}{n-1}}}{N! L^{2\ell+n}}\dfrac{(p!)^2}{(2p)!(2p+1)!}\nonumber\\
&&\times\dfrac{\Gamma\left(N+\frac{n+1}{2}+\ell+\sqrt{m_0^2L^2+(\frac{n+1}{2})^2}\right)}{\Gamma\left(N+1+\sqrt{m_0^2L^2+(\frac{n+1}{2})^2}\right)}\dfrac{1}{\Gamma\left(\ell+\frac{n-1}{2}\right)}\nonumber\\
&&\times \prod_{k=1}^N\left(\ell+\frac{n-1}{2}+k\right) \prod_{k^{\prime}=1}^p (k^{\prime 2}+4\bar{\omega}^2)\;.\label{case1}
\end{eqnarray}
This equation shares a similar structure to the corresponding result in $D=4$. From the definition of $\bar{\omega}$ in Eq.~\eqref{superradiance}, we find that in the superradiant regime, $\bar{\omega}<0$ which implies $\delta>0$. In this superradiant regime the wave function of the scalar field will grow with time which means the BH is unstable. Moreover, from Eq.~\eqref{superradiance}, one may get a condition for the onset of the superradiant instability, i.e.
\begin{eqnarray}
\dfrac{q}{q_c} > \sqrt{\frac{2(n-1)}{n}} \dfrac{\omega_N}{e}\;,\label{onset1}
\end{eqnarray}
where $\omega_N$ is given in Eq.~\eqref{AdSfrequency}. For a massless field, Eq.~\eqref{onset1} simplifies to
\begin{eqnarray}
\dfrac{q}{q_c} > \sqrt{\frac{2(n-1)}{n}} \dfrac{2N+n+1+\ell}{e L}\;.\label{onset2}
\end{eqnarray}

\subsection{$\ell$ is not an integer multiple of $(n-1)$}
For this case, the corresponding Gamma function in Eq.~\eqref{imaginarypart} can be simplified as
\begin{equation}
\dfrac{\Gamma(1-2\alpha)\Gamma(\alpha)}{\Gamma(2\alpha-1)\Gamma(1-\alpha)}=-\dfrac{1}{2\cos\frac{\pi \ell}{n-1}}\dfrac{\Gamma^2(1+\frac{\ell}{n-1})}{\Gamma(1+\frac{2\ell}{n-1})\Gamma(2+\frac{2\ell}{n-1})}\;.\nonumber
\end{equation}
If $\dfrac{\ell}{n-1}\neq p+\frac{1}{2}$, then cos$\frac{\pi\ell}{n-1}\neq0$, and the parameter $\delta$ becomes complex (not simply real as in the previous case).
In this case the real part of $\delta$ reflects the instability, which is given by
\begin{eqnarray}
\mbox{Re} \delta&=&-2\bar{\omega}\dfrac{(r_+^{n-1}-r_-^{n-1})^{1+\frac{2\ell}{n-1}}}{N! L^{2\ell+n}}\prod_{k=1}^N(\ell+\frac{n-1}{2}+k)\times\nonumber\\
&&\times \dfrac{\Gamma^4(1+\frac{\ell}{n-1})}{\Gamma(1+\frac{2\ell}{n-1})\Gamma(2+\frac{2\ell}{n-1})\Gamma(\ell+\frac{n-1}{2})}\times\nonumber\\
&&\times\dfrac{\Gamma(N+\frac{n+1}{2}+\ell+\sqrt{m_0^2L^2+(\frac{n+1}{2})^2})}{\Gamma(N+1+\sqrt{m_0^2L^2+(\frac{n+1}{2})^2})}\;,\label{case2}
\end{eqnarray}
where we have expanded the terms $\Gamma(x-2i\bar{\omega})$ around small $\bar{\omega}$ to clearly distinguish the superradiant regime. Thus, when $\bar{\omega}<0$, we obtain Re$\delta$ $>0$ which implies the BH is also unstable, and the corresponding onset of such instability is governed by Eqs.~\eqref{onset1} for massive field and \eqref{onset2} for massless field.

If $\frac{\ell}{n-1}=p+\frac{1}{2}$, the matching method fails; a similar situation occurs for extremal Kerr BHs~\cite{Rosa:2009ei}\footnote{We thank Jo\~ao Rosa for drawing our attention to this point.}. In order to make this point clear, we can do the following analysis. Firstly, from the definition of $\alpha$ in Eq.~\eqref{alpha} and the condition $\frac{\ell}{n-1}=p+\frac{1}{2}$, one observes that the first expansion term inside the bracket of  Eq.~\eqref{NearsolFar} is divergent, which means that we cannot expand Eq.~\eqref{NearSol} into Eq.~\eqref{NearsolFar} anymore when $\frac{\ell}{n-1}=p+\frac{1}{2}$. Alternatively, using a property of hypergeometric function~\cite{abramowitz+stegun}, we shall expand Eq.~\eqref{NearSol} as
\begin{eqnarray}
{\rm R} \sim && \Gamma(1-2i\bar{\omega})\left[-\dfrac{(r_+^{n-1} - r_-^{n-1})^\alpha \zeta}{\Gamma(1-\alpha)\Gamma(1-\alpha-2i\bar{\omega})\Gamma(2\alpha)}\dfrac{1}{r^{n-1+\ell}} \right.\nonumber \\ && \left.
+\dfrac{\Gamma(2\alpha-1)(r_+^{n-1}-r_-^{n-1})^{1-\alpha}}{\Gamma(\alpha)\Gamma(\alpha-2i\bar{\omega})}r^{\ell}\right]\;,\label{NearsolFar2}
\end{eqnarray}
with
\begin{equation}
\zeta=\log\left(\dfrac{r_+^{n-1}-r_-^{n-1}}{r^{n-1}}\right)+\gamma+\psi(\alpha)-\psi(2\alpha)+\psi(\alpha-2i\bar{\omega})\;,\nonumber
\end{equation}
where $\gamma$ is Euler constant and $\psi(x)$ denotes the Digamma function.
Because the $\log r$ term is associated with distinct powers of $r$, it is impossible to match Eqs.~\eqref{farsolution@near} and \eqref{NearsolFar2}. For this case, we have to appeal to a numerical solution which is discussed in the next section.

\section{Numerical results}
\label{numerical}
In order to confirm the above analytical results and calculate the quasinormal frequency for the special case $\frac{\ell}{n-1}=p+\frac{1}{2}$ where the analytical method fails, we shall solve, in this section, Eq.~\eqref{RadialEq2} numerically. We use a direct numerical integration method to obtain the quasinormal frequency of the BH. To do so, taking the boundary conditions near the horizon in Eq.~\eqref{boundaryingoing} and at infinity in Eq.~\eqref{boundarydecaying}, we expand the radial function near the horizon as
\begin{equation}
{\rm X} \sim e^{-i(\omega-\omega_0)r_\ast} \sum_{j=0}^{\infty}\alpha_j(r-r_+)^j\;,\label{nearHExp}
\end{equation}
and at infinity as
\begin{equation}
{\rm X} \sim r^{-\frac{1}{2}(1+\sqrt{4m_0^2L^2+(n+1)^2})} \sum_{j=0}^{\infty}\frac{\beta_j}{r^j}\;.\label{infExp}
\end{equation}
The series expansion coefficients can be derived directly after inserting these expansions into Eq.~(\ref{RadialEq2}).
We use the series expansion near the horizon Eq.~\eqref{nearHExp} to initialize the radial system Eq.~\eqref{RadialEq2} from a point $r_s$ which is close to $r_+$ through the relation $r_s=(1+0.01)r_+$, and integrate the radial system outwards up to a radial value $r_m$. Similarly we can also use Eq.~\eqref{infExp} as initial condition to integrate the radial system inward from $r_l=1000r_+$ down to $r_m$. Then we have two solutions at an intermediate radius $r_m$, and these two solutions are linearly dependent if their Wronskian vanishes at $r_m$. Using a secant method one can solve $W(\omega,r_m)=0$ iteratively to look for the quasinormal frequency of the BH. We also varied $r_s$, $r_m$ and $r_l$ to check the numerical accuracy.

We list some numerical results in TABLES I-V. Note that all physical quantities are normalized by the AdS radius $L$ and we set $L=1$. In the first three tables, we focus on the fundamental modes of massless fields because they are typically the most unstable modes. To check the mass effect on the validity of the analytical formulas, we also consider $m_0=0.5$ and $m_0=3.0$ in the last two tables. As a check on our numerical method, we have calculated the quasinormal frequencies for small Schwarzschild-AdS BH and we obtained results which are in good agreement with those reported in~\cite{Konoplya:2002zu}.

In order to address the special case for which the analytical method failed, i.e when $\ell=(p+\frac{1}{2})(n-1)$, we chose $n=3$ (five dimensional spacetime) and $\ell=1$, corresponding to $p=0$ in our condition. The result is shown in TABLE I, with $r_+=0.1$, field mass $m_0=0$ and field charge $e=8$. It shows clearly that superradiant instability appears when $q/q_c$ satisfies the condition~\eqref{onset2}. Moreover, we also list numerical results for $\ell=0$ mode in TABLE I. It shows that frequencies of the odd modes ($\ell=1$) and even modes ($\ell=0$) have similar behavior; in other words, there is nothing special for odd modes.

To confirm the validity of the analytical quasinormal frequency formulas in Eqs.~\eqref{case1} and \eqref{case2}, we also compare some analytical results with numerical data in TABLES~II-V. In TABLE II, we present analytical results obtained from Eq.~\eqref{case1} and numerical results with $r_+=0.01$, $e=6$ for the $\ell=0$ massless fundamental mode in five dimensional spacetimes. They show good agreement; the difference is smaller than 1\%. In TABLE III, we present analytical results obtained from Eq.~\eqref{case2} and numerical data, for the $\ell=1$ fundamental mode with $r_+=0.01$, $e=10$, $m_0=0$ in $D=6$, and they show good agreements as well. From these two tables, we confirm the validity of analytical matching method for $m_0=0$. Results for non-zero mass are reported in TABLES~IV-V. Two conclusions may be drawn from these tables. Firstly, as the mass is increased the agreement between analytic and numerical method becomes worse. This is expected in view of the approximation discussed in Sec.~\ref{secmatching}. Secondly, as the mass increases, the mode with $q/q_c=0.9$ becomes stable. This is in agreement with Eq.~\eqref{onset1}, since for the parameters in TABLE~V, superradiance is only expected for $q/q_c \gtrsim 1.1$.

\begin{table}
\label{3DRNAdS}
\caption{Frequencies of the fundamental modes with different $\ell$ for a BH with $r_+=0.1$, $e=8$, $m_0=0$ in $D=5$.}
\begin{ruledtabular}
\begin{tabular}{ l l l }
$q/q_c$ & $\ell$=0 & $\ell$=1 \\
\hline\\
0.1 & 3.958 - 1.335$\times 10^{-2}$ i & 4.978 - 2.689$\times 10^{-4}$ i\\
0.3 & 3.997 - 6.435$\times 10^{-3}$ i & 4.998 - 1.367$\times 10^{-4}$ i\\
0.5 & 4.030 - 1.522$\times 10^{-3}$ i & 5.014 - 5.053$\times 10^{-5}$ i\\
0.7 & 4.058 + 1.996$\times 10^{-3}$ i & 5.028 - 2.596$\times 10^{-6}$ i\\
0.8 & 4.070 + 3.198$\times 10^{-3}$ i & 5.034 + 7.524$\times 10^{-6}$ i\\
0.9 & 4.081 + 3.954$\times 10^{-3}$ i & 5.040 + 9.597$\times 10^{-6}$ i\\
\end{tabular}
\end{ruledtabular}
\end{table}

\begin{table}
\label{comp1}
\caption{Comparison of the frequencies for the $\ell=0$ fundamental modes of a BH with $r_+=0.01$, $e=6$, $m_0=0$ in $D=5$.}
\begin{ruledtabular}
\begin{tabular}{ l l l }
$q/q_c$ & Im$(\omega)$ (numerical) & Im$(\omega)$ (analytical) \\
\hline\\
0.1 & -1.053$\times 10^{-5}$  & -1.0441$\times 10^{-5}$ \\
0.3 & -7.369$\times 10^{-6}$  & -7.3230$\times 10^{-6}$ \\
0.5 & -4.222$\times 10^{-6}$  & -4.2050$\times 10^{-6}$ \\
0.7 & -1.088$\times 10^{-6}$  & -1.0870$\times 10^{-6}$ \\
0.9 & 2.023$\times 10^{-6}$  & 2.0310$\times 10^{-6}$ \\
\end{tabular}
\end{ruledtabular}
\end{table}

\begin{table}
\label{comp2}
\caption{Comparison of the frequencies for the $\ell=1$ fundamental modes of a BH with $r_+=0.01$, $e=10$, $m_0=0$ in $D=6$.}
\begin{ruledtabular}
\begin{tabular}{ l l l }
$q/q_c$ & Im$(\omega)$ (numerical) & Im$(\omega)$ (analytical) \\
\hline\\
0.1 & -4.377$\times 10^{-11}$  & -4.3678$\times 10^{-11}$ \\
0.3 & -2.830$\times 10^{-11}$  & -2.8274$\times 10^{-11}$ \\
0.5 & -1.342$\times 10^{-11}$  & -1.3418$\times 10^{-11}$ \\
0.7 & -1.538$\times 10^{-12}$  & -1.5371$\times 10^{-12}$ \\
0.8 & 2.283$\times 10^{-12}$  & 2.2846$\times 10^{-12}$ \\
0.9 & 3.778$\times 10^{-12}$  & 3.7844$\times 10^{-12}$ \\
\end{tabular}
\end{ruledtabular}
\end{table}

\begin{table}
\label{compmass1}
\caption{Comparison of the frequencies for the $\ell=0$ fundamental modes of a BH with $r_+=0.01$, $e=6$, $m_0=0.5$ in $D=5$.}
\begin{ruledtabular}
\begin{tabular}{ l l l }
$q/q_c$ & Im$(\omega)$ (numerical) & Im$(\omega)$ (analytical) \\
\hline\\
0.1 & -1.093$\times 10^{-5}$  & -1.0844$\times 10^{-5}$ \\
0.3 & -7.711$\times 10^{-6}$  & -7.6617$\times 10^{-6}$ \\
0.5 & -4.498$\times 10^{-6}$  & -4.4797$\times 10^{-6}$ \\
0.7 & -1.300$\times 10^{-6}$  & -1.2977$\times 10^{-6}$ \\
0.9 &  1.878$\times 10^{-6}$  & 1.8842$\times 10^{-6}$ \\
\end{tabular}
\end{ruledtabular}
\end{table}

\begin{table}
\label{compmass2}
\caption{Comparison of the frequencies for the $\ell=0$ fundamental modes of a BH with $r_+=0.01$, $e=6$, $m_0=3.0$ in $D=5$.}
\begin{ruledtabular}
\begin{tabular}{ l l l }
$q/q_c$ & Im$(\omega)$ (numerical) & Im$(\omega)$ (analytical) \\
\hline\\
0.1 & -2.379$\times 10^{-5}$  & -2.3423$\times 10^{-5}$ \\
0.3 & -1.886$\times 10^{-5}$  & -1.8637$\times 10^{-5}$ \\
0.5 & -1.399$\times 10^{-5}$  & -1.3850$\times 10^{-5}$ \\
0.7 & -9.155$\times 10^{-6}$  & -9.0632$\times 10^{-6}$ \\
0.9 & -4.321$\times 10^{-6}$  & -4.2765$\times 10^{-6}$ \\
\end{tabular}
\end{ruledtabular}
\end{table}

\section{Discussion and Final Remarks}
\label{discussion}
In this paper we have studied  the superradiant instability of small charged AdS BHs in $D$ dimensions, in the presence of a charged scalar field. Such a systematic study for arbitrary $D$ was absent in the literature where only results for $D=4,5$ had been considered~\cite{Uchikata:2011zz,Aliev:2008yk,Dias:2011tj}.

Firstly, we solved the Klein-Gordon equation for a charged scalar field in charged AdS BHs with a standard matching method. We found that the relation between the angular momentum quantum number $\ell$ and the spacetime dimension $D$ plays an important role in determining the analytical quasinormal frequency formula. When $\ell=p(D-3)$, for a non-negative integer $p$, we found that the quasinormal frequencies of the small RN-AdS BHs have only an imaginary correction to the AdS normal frequencies. This is the case for all modes (i.e. all $\ell$) in $D=4$, even $\ell$ in $D=5$, $\ell=0,3,6,9,\dots$ in $D=6$, $\ell=0,4,8,12,\dots$ in $D=7$ and so on.

A more subtle case occurs when $\ell=(p+\frac{1}{2})(D-3)$. For this case the matching method is inapplicable because a $\log r$ term appears in the near region solution -~Eq.~\eqref{NearsolFar2} - which cannot be matched to Eq.~\eqref{farsolution@near}. Failure to observe this limitation has led to a claim that odd $\ell$ modes in $D=5$ did not exhibit superradiance~\cite{Aliev:2008yk}. Here we have shown otherwise that superradiant instability indeed exists by a numerical investigation which is mandatory for analysing this case, in view of the invalidity of the matching method. A similar conclusion should apply to all cases defined by  $\ell=(p+\frac{1}{2})(D-3)$, i.e, odd $\ell$ in $D=5$, $\ell=2,6,10,14,\dots$ in $D=7$ and so on. Observe that this case can only occur in odd dimensions.

Finally, all other cases have a complex correction to the AdS normal frequencies, i.e. the real part of the frequency is also shifted.

Our analytic results show good agreement with the numerical results in Sec. \ref{numerical}. In particular a central conclusion is that all $\ell$ modes in all dimensions, for sufficiently large field charge $e$ display superradiance. In particular, in $D=4$, the dependence of the instability on the various parameters seems to be in qualitative agreement with the study of cavity BHs in $D=4$ \cite{Herdeiro:2013pia,Hod:2013fvl,Degollado:2013bha} and it would be interesting to make a more detailed comparison between the two cases.

Let us close with two questions. Firstly, is there a simple pattern for the behaviour of the frequencies as $D\rightarrow \infty$? A preliminary analysis could not unveil a simple formula. Finding such behaviour would be relevant in view of the recent interest on General Relativity in the large $D$ limit \cite{Emparan:2013moa}. Secondly, can one follow this instability numerically into the non-linear regime? It is expected that the end-point of the instability will be a hairy charged AdS BH (see e.g \cite{Dias:2011tj}), but it remains to show explicitly, by following the time evolution of the unstable scalar-gravity system, that it is indeed so.

\bigskip

\noindent{\bf{\em Acknowledgements.}}
It is a pleasure to thank Marco Sampaio, Jo\~ao Rosa and J.C. Degollado for discussions and suggestions. M.W. is funded by FCT through the grant SFRH/BD/51648/2011. The work in this paper is also supported by the grants PTDC/FIS/116625/2010 and  NRHEP--295189-FP7-PEOPLE-2011-IRSES.

\bibliographystyle{h-physrev4}
\bibliography{BHbomb}


\end{document}